\documentclass[twocolumn,preprintnumbers,amsmath,amssymb,floatfix,prd]{revtex4-2}
\usepackage{epsfig}
\usepackage{dcolumn}
\begin{document}
\title{Pion Nuclear Fragmentation Functions Revisited}
\author{Matias Doradau}
\email{matiasdoradau@outlook.com} 
\affiliation{Universidad de Buenos Aires, Facultad de Ciencias Exactas 
y Naturales, Departamento de F\'{\i}sica and IFIBA-CONICET, Ciudad Universitaria, (1428) Buenos Aires, Argentina}
\author{Ramiro Tomas Martinez}
\email{rmartinez@df.uba.ar} 
\affiliation{Universidad de Buenos Aires, Facultad de Ciencias Exactas 
y Naturales, Departamento de F\'{\i}sica and IFIBA-CONICET, Ciudad Universitaria, (1428) Buenos Aires, Argentina}
\author{Rodolfo Sassot}
\email{sassot@df.uba.ar} 
\affiliation{Universidad de Buenos Aires, Facultad de Ciencias Exactas 
y Naturales, Departamento de F\'{\i}sica and IFIBA-CONICET, Ciudad Universitaria, (1428) Buenos Aires, Argentina}
\author{Marco Stratmann}
\email{marco.stratmann@uni-tuebingen.de}
\affiliation{Institute for Theoretical Physics, University of T\"ubingen, Auf der 
Morgenstelle 14, 72076 T\"ubingen, Germany}
\begin{abstract}
We revisit the notion of nuclear parton-to-pion fragmentation functions 
at next-to-leading order accuracy as an effective description of hadroproduction 
in nuclear environments such as in semi-inclusive lepton-nucleus deep-inelastic scattering and 
in single inclusive proton-nucleus collisions. 
We assess their viability in the face of very precise data collected for the latter 
at the CERN-LHC over the past decade
as well as recent measurements of the former carried out by the CLAS experiment at JLab.   
\end{abstract}
%
\maketitle
\section{Introduction}
Parton distributions functions (PDFs) and fragmentation functions (FFs) are key ingredients in the 
perturbative Quantum Chromodynamics (QCD) description of hard scattering processes involving nucleons 
in the initial state and identified hadrons emerging from hard partonic interactions \cite{Feynman:2019rot}. 
The former encode the information on the partonic content of nucleons entering the hard interaction 
and the latter that on the hadronization process. 
Both involve distance scales larger than the QCD scale $\Lambda_{QCD}$ and, hence, are beyond the reach 
of perturbative methods in QCD. 

Nevertheless, PDFs and FFs can be successfully determined from global QCD analyses, combining data and 
relevant short distance cross sections computed order by order in perturbation theory, 
assuming they factorize \cite{Collins:1989gx} in the presence of a large enough energy scale.
Over the past decades these analyses have progressed significantly in terms of precision and sophistication, 
involving perturbative calculations at next-to-next-to-leading order (NNLO) accuracy and beyond 
as well as an ever increasing amount and variety of very precise data \cite{Hou:2019efy,Bailey:2020ooq,NNPDF:2021njg}.

In deep-inelastic scattering (DIS) and Drell-Yan (DY) processes involving a nucleus instead of free nucleons, 
the influence of the nuclear environment can be effectively factorized into modified nuclear parton distribution functions (nPDFs). 
These nPDFs depend on energy through the standard scale evolution equations of QCD, 
and conventional partonic hard scattering cross sections can be utilized in calculations.
Within the precision of the available data, this approach was demonstrated to yield an excellent approximation 
at next-to-leading order (NLO) accuracy 
of perturbative QCD already twenty years ago \cite{deFlorian:2003qf}, and, more recently, has been extended to NNLO 
in a couple of analyses, see, e.g.\ \cite{Klasen:2023uqj} for a review.

In spite of their phenomenological success for fully inclusive processes, nPDFs are unable to reproduce 
the significant differences found when comparing production processes of specific identified hadron species 
occurring in a nuclear environment with large atomic mass $A$ to similar measurements 
involving only light nuclei or proton targets \cite{ref:oldexp,HERMES:2009uge}. 
Even though the observed differences induced by the nuclear media can be attributed to a variety of conceivable mechanisms
besides the well-known modification of parton densities in nuclei \cite{Arleo:2008dn}, in reference \cite{Sassot:2009sh} 
it was conjectured that QCD factorization could be extended to incorporate final state nuclear effects by
introducing {\em effective} nuclear fragmentation functions (nFFs) depending on $A$.
It was demonstrated that such an approximation worked remarkably well in practice within the precision of the data 
available at that time. The idea was explored further in \cite{Zurita:2021kli}.

Since then, several new sets of measurements have been made available, calling for a more exhaustive, critical assessment of nFFs.
First and foremost, the CERN-Large Hadron Collider (LHC) has produced remarkably precise data on hadroproduction 
with both proton-proton and proton-lead beams in a vast range of values of the hadron's transverse momentum $p_T$ and 
pseudo-rapidity $\eta$ \cite{ALICE:2016dei,ALICE:2021est,LHCb:2022tjh}, 
very much extending previously available data from BNL-RHIC \cite{STAR:2003oii,PHENIX:2006mhb} hitherto analyzed.
In particular, the recent LHC data from proton-proton collisions allowed for a much more precise extraction 
of vacuum FFs \cite{Borsa:2021ran,Borsa:2023zxk,PhysRevD.110.114019,PhysRevLett.132.261903}, which, at variance with the FFs sets \cite{deFlorian:2007aj} 
used in the original analysis of nFFs in Ref.~\cite{Sassot:2009sh}, reproduce fairly accurately 
the cross sections measured up to the highest center-of-mass system (c.m.s.) energies $\sqrt{S}$.
In turn, the data taken in proton-lead (pPb) collisions allow for a more stringent extraction of nFFs. 
Most importantly, measurements by LHCb at both forward and backward rapidity ranges \cite{LHCb:2022tjh} probe nPDFs 
at different parton momentum fractions and, as we shall see, also help to discriminate 
regions dominated by either quark or gluon fragmentation in hadron production. Additionally, transverse momentum dependent (TMD) fragmentation function have been addressed in \cite{PhysRevLett.129.242001,Alrashed:2023xsv}.

On the other hand, the CLAS collaboration at Thomas Jefferson National Accelerator Facility (JLab) 
has produced very precise nuclear semi-inclusive DIS (SIDIS) multiplicity ratios 
for identified pions off C, Fe, and Pb targets, respectively \cite{CLAS:2021jhm}. 
The data are taken at a kinematic range that is complementary to the one covered already
by the preceding DESY-HERMES experiment \cite{HERMES:2009uge}
used in the analysis of Ref.~\cite{Sassot:2009sh}. They also include different nuclei to better map the $A$-dependence of nFFs.

In addition, the sets of nPDFs, which were used as an input to our original analysis of nFFs, have evolved considerably. 
While the sets available back then were mostly based on fixed target nuclear DIS data, the most recent extractions of nPDFs 
include a variety of observables ranging from  Drell-Yan $W^{\pm}$ and $Z$-boson production to 
high-$p_T$ prompt photon and jet production 
in proton-nucleus collisions \cite{Klasen:2023uqj}. 

Finally, on the methodological side of global QCD analyses, there have been many crucial developments
recently, mainly regarding reliable estimates of the uncertainties inherent to extractions of PDFs and FFs from data. 
Specifically, we will estimate the uncertainties on the resulting nFFs systematically 
by means of a Monte Carlo error sampling method, similar to the one used for corresponding
extractions vacuum FFs \cite{Borsa:2021ran,Borsa:2023zxk}. These recent sets of FFs will also serve us
as a baseline to quantify a potential suppression or enhancement in parton-to-hadron fragmentation
in a nuclear environment.

Therefore, it is very timely to review the notion of medium modified or nuclear FFs introduced in Ref.~\cite{Sassot:2009sh}
and to critically reassess the range of validity of the assumed underlying effective factorization.
In this paper, we will revisit and substantially extend the analysis originally performed in \cite{Sassot:2009sh} 
by including the latest experimental information from the ALICE, LHCb, and CLAS collaborations as well as
up-to-date sets of vacuum FFs, PDF, and nPDFs as inputs, and by providing careful estimates of the 
uncertainties of the resulting nFFs based on Monte Carlo replicas.
We shall demonstrate that the new data still favor an overall picture featuring
a strong suppression of quark fragmentation and a moderate enhancement 
of hadrons originating from gluons, both effects growing with nuclear size $A$. 
For hadroproduction in proton-lead collisions, the results will dependent also on the choice of 
the nPDFs set, and the best agreement with data will be found with the nNNPDF3.0 set \cite{AbdulKhalek:2022fyi}. 
We note, that we will adopt the most recent ALICE results for pion production at very large transverse momentum \cite{ALICE:2021est}  
not in the fit, as we shall explain in detail, but to benchmark our newly extracted set of effective 
nFFs and to delineate its possible limitations.

The remainder of the paper is organized as follows:
in the next section, we focus on the main new ingredients that we are going to utilize 
in our global analysis of nFFs, 
namely the recent data sets from CERN and JLab and their respective kinematical coverage.
We introduce the nuclear observables/ratios relevant for the extraction of nFF: 
$R^{\pi}_{p(d)A}$ and $R^{\pi}_{A}$ for the production of pions in proton(deuteron)-nucleus collisions 
and lepton-nucleus SIDIS, respectively. 
We also show results for two different recent sets of nPDFs, nNNPDF3.0 \cite{AbdulKhalek:2022fyi}
and TUJU21 \cite{Helenius:2021tof},
that we take as representatives of the current range of uncertainties in the extraction of nPDFs from data.
In Section III, we present the outcome of our new global QCD analysis in terms of 
the resulting effective medium modified FFs,
assess the degree of agreement with the data, 
and discuss possible limitations of the framework
by using the latest ALICE data as a benchmark.
Finally, we give our conclusions and present a brief outlook.

\section{Review of new Ingredients}
%
\begin{figure*}[t!]
\epsfig{figure=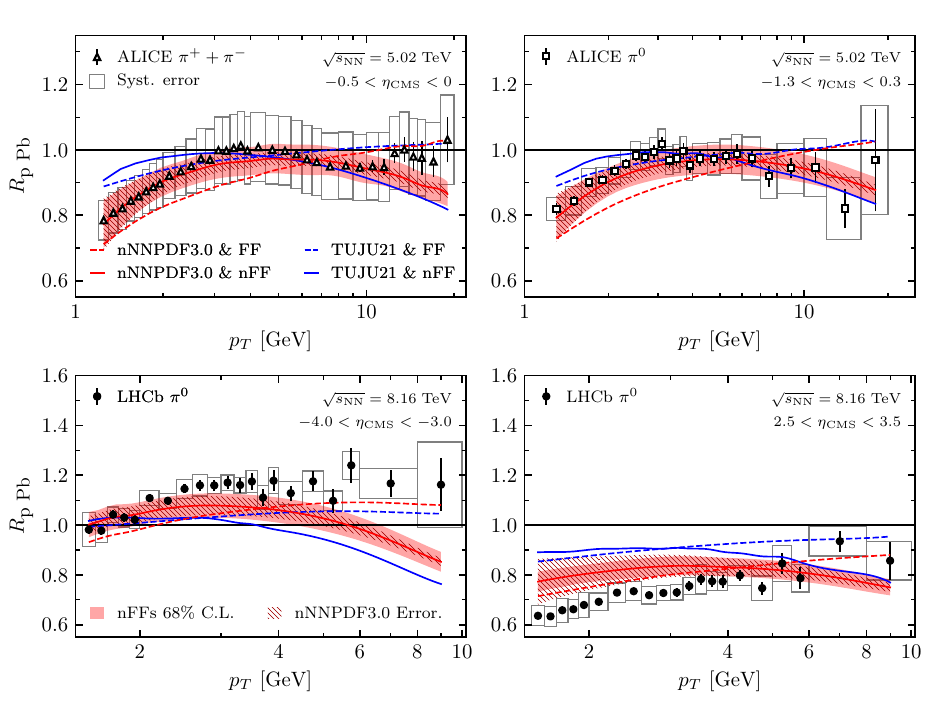,width=0.9\textwidth}
\vspace*{-0.4cm}
\caption{Nuclear modification factors from ALICE \cite{ALICE:2016dei} (upper panels)
and LHCb \cite{LHCb:2022tjh} (lower panels)
for charged and neutral pions at 
$\sqrt{S}=5.02$ and $8.16\,\mathrm{TeV}$, respectively, for different rapidity ranges
compared to calculations with (solid lines) and without (dashed lines) 
including final state nuclear effects; see text. Two different sets of nPDFs
(nNNPDF3.0 \cite{AbdulKhalek:2022fyi} and TUJU21 \cite{Helenius:2021tof}) are used.
The shaded bands illustrate the uncertainties of the nPDFs by nNNPDF3.0 (hatched) and for
our newly extracted set of nFFs (solid).
}\label{fig1}
\end{figure*}
Both the ALICE and the LHCb collaborations at the LHC have produced remarkably precise data 
on charged and neutral pion production in proton-lead (pPb) collisions at c.m.s.\ energies
$\sqrt{S}=5.02$ and $8.16\,\mathrm{TeV}$ and
in different ranges of $p_T$ and $\eta$ \cite{ALICE:2016dei,ALICE:2021est,LHCb:2022tjh}. 

The data are usually presented and discussed in terms of nuclear modification factors $R^{\pi}_{p(d)A}$, defined as 
ratios of the measured differential cross sections for collisions involving a heavy nucleus, 
normalized to the number of participating nucleons $A$, and the corresponding proton-proton (pp) reference.
Specifically, in case of high-$p_T$ pion production it reads
\begin{equation}
R^{\pi}_{p(d)A} \equiv \frac{1}{A}\,\frac{d^2\sigma^{\pi}_{p(d)A}/dp_T d\eta}
{d^2\sigma^{\pi}_{pp}/dp_T d\eta}\;.
\label{eq:rhicrate}
\end{equation}
We note that $d^2\sigma^{\pi}_{pp}$ is often being obtained by interpolating existing experimental
results to the desired values of $\sqrt{S}$ of the $p(d)A$ measurement \cite{ALICE:2016dei,ALICE:2021est,LHCb:2022tjh}.

Theoretical estimates for both the numerator and the denominator of (\ref{eq:rhicrate})
can be computed consistently, order by order in perturbation theory
and with any desired choice of sets of PDFs, nPDFs, and FFs, 
by invoking, as usual, the factorization theorem:
\begin{eqnarray}
\nonumber
\frac{d^2\sigma^{\pi}}{dp_T d\eta} &=&
\sum_{a,b,c} f_a(x_a,\mu_f) \otimes f_b(x_b,\mu_f)
\otimes D_c^{\pi}(z_c,\mu_{f^{\prime}})  \\
&&
\otimes \;d\hat{\sigma}_{ab\to cX}(S,\alpha_s,x_a,x_b,z_c,\mu_f,\mu_r,\mu_{f^\prime})\;.
\label{eq:xsec}
\end{eqnarray}
Here, the sum is over all contributing partonic short-distance
scattering cross sections $d\hat{\sigma}_{ab\to cX}$,
which are currently known up to NLO accuracy in case of hadroproduction \cite{Aversa:1988vb}.
This also limits the accuracy at which a global QCD analysis 
of nFFs can be performed for the time being. 
The scales $\mu_f$ and $\mu_{f^{\prime}}$ are introduced to factorize initial and final state collinear singularities 
into the scale dependent PDFs and FFs, $f_{a,b}(x_{a,b},\mu_f)$ and $D_c^h(z_c,\mu_{f^{\prime}})$,
respectively, where $x_{a,b}$ and $z_c$ denote the usual collinear momentum fractions.
$\mu_r$ denotes the energy scale at which the strong coupling $\alpha_s$ is being renormalized.

The nuclear modification factor $R^{\pi}_{p(d)A}$ in Eq.~(\ref{eq:rhicrate}) has the advantage that
the rather sizable dependence of the cross sections on the actual choice for the factorization scales 
$\mu_f,\mu_{f^{\prime}}=\mathcal{O}(p_T)$ at NLO accuracy tends to cancel to a large extent, as is the case
with other potential sources of experimental and theoretical ambiguities, mostly 
related to the absolute normalization of the cross sections. 

We shall notice, that our baseline set of NLO vacuum FFs, BDSS21 \cite{Borsa:2021ran,Borsa:2023zxk},
was extracted from the same precise $pp$ data from the LHC entering the experimental
determination of the nuclear modification factor (\ref{eq:rhicrate}) 
except for the latest ALICE data set \cite{ALICE:2021est}, which extends measurements to larger values 
of $p_T $ of up to about $200\,\mathrm{GeV}$.
The latter data were not available in time of the BDSS21 analysis. 
Hence, we will use this data set only to benchmark our new extraction of the nFFs, see Section III.

\begin{figure*}[th!]
\centering
\epsfig{figure=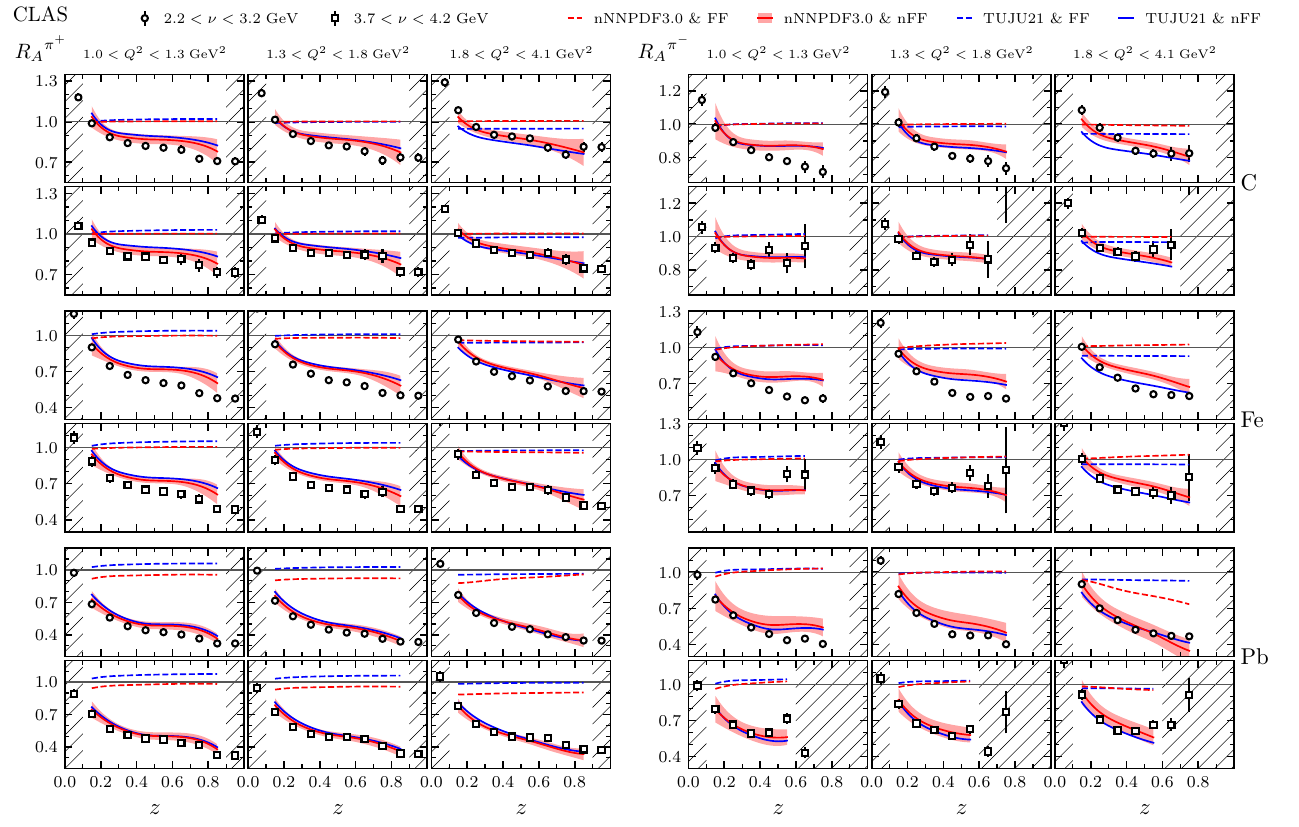,width=\textwidth}
\vspace*{-0.4cm}
\caption{Nuclear modification factors defined in Eq. (\ref{eq:mult}) for SIDIS multiplicities for
positively (left panel) and negatively (right panel) charged pions 
for various kinematic bins and nuclei from CLAS \cite{CLAS:2021jhm}
compared to calculations with (solid lines) and without (dashed lines) 
including final state nuclear effects; see text.
The shaded bands illustrate the uncertainties of 
our newly extracted set of nFFs.}
\label{fig2}
\end{figure*}

\begin{figure*}[thb!]
\hspace*{-1.9cm}
\epsfig{figure=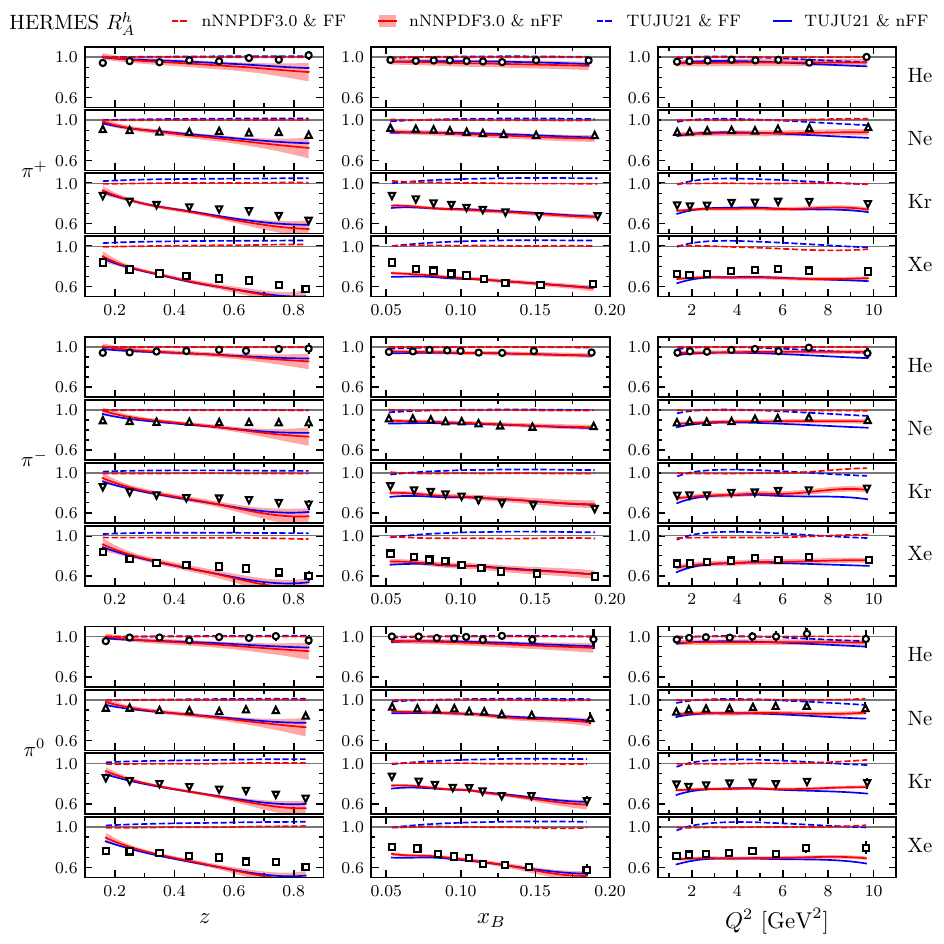,width=0.8\textwidth}
\vspace*{-0.2cm}
\caption{Same as in Fig.~\ref{fig2} but now for the nuclear modification factors
for SIDIS multiplicities for charged and neutral pions from HERMES \cite{HERMES:2009uge}.}
\label{fig3}
\end{figure*}

To set the stage, we will focus in the remainder of this section on how well data are described 
by first neglecting any final state nuclear effects, i.e., on what can be achieved by combining 
nPDFs with vacuum FFs in Eq.~(\ref{eq:xsec}) to compute the numerator of the nuclear 
modification factor (\ref{eq:rhicrate}).
But to facilitate the presentation, Figs.~\ref{fig1}-\ref{fig3} show the results computed
with our newly fitted effective nFFs as well (solid lines), but we shall postpone a discussion of final state 
nuclear effects and the details of our global analysis to Section III.

We start in Fig.~\ref{fig1} by showing nuclear modification factors for charged and neutral pion production
in pPb collisions at $\sqrt{S}=5.02\,\mathrm{TeV}$ and $\sqrt{S}=8.16\,\mathrm{TeV}$ measured by 
the ALICE (upper panels) and LHCb (lower panels) collaboration, respectively.
The results by ALICE are obtained close to mid rapidity and therefore mainly
probe parton momentum fractions smaller than $x\approx 0.1$, i.e., roughly the transition
from the shadowing to the anti-shadowing domain of nPDFs 
as the pion's transverse momentum increases.
The underlying kinematics of the data is such that the observed pions stem predominantly 
from hadronizing gluons. 
Even as $p_T$ increases, the rising contribution from quark fragmentation will
still remain smaller than the one from gluons.
This explains the particular importance of the ALICE data in pinning down any
possible medium induced effects in the gluon fragmentation in our analysis.
We note that a detailed discussion of the correlation between the kinematical variables 
and the relevance of different partonic channels $\;d\hat{\sigma}_{ab\to cX}$
in Eq.~(\ref{eq:xsec}) at typical LHC c.m.s.\ energies can be found in
Ref.~\cite{Sassot:2010bh}, see, in particular, Fig.~6.

The dashed lines in Fig.~\ref{fig1} 
show theoretical estimates of the nuclear modification factors in Eq.~(\ref{eq:rhicrate}), 
computed at NLO accuracy with two different sets of nPDFs,
nNNPDF3.0 \cite{AbdulKhalek:2022fyi} (red)
and TUJU21 \cite{Helenius:2021tof} (blue), respectively, and 
the BDSS21 set of vacuum FFs \cite{Borsa:2021ran,Borsa:2023zxk},
i.e., assuming {\em no} nuclear effects during the hadronization process.
We have picked these two sets of nPDFs for our purposes since neither of them makes use of
any pion production pA or SIDIS data in their determination. 
This is crucial for the consistency of our analysis, since we want to
separate potential final state nuclear effects in pA data
from those attributed to the initial state, i.e., nPDFs.

As can be seen, both results based on vacuum FFs roughly reproduce the trend of the ALICE data
within the experimental uncertainties once the large systematic errors are taken into account.
The shape of the curves with increasing $p_T$ 
reflects the amount of shadowing and antishadowing encoded in each set of nPDFs. 
Typically, both calculations fall within a range less than 10\% away from the 
central values of the data,
which hints towards a maximum limit for final state nuclear effects at
mid rapidity for LHC c.m.s.\ energies.
On the other hand, the differences between the two estimates are
fairly representative of the nPDFs uncertainties at play here,
which turn out roughly twice as large as the maximum size of 
potential final state nuclear effects; see the hatched bands in Fig~\ref{fig1} as
an illustration of nPDF uncertainties.

The two lower panels of Fig.~\ref{fig1} show neutral pion data at
$\sqrt{S}=8.16\,\mathrm{TeV}$ from LHCb \cite{LHCb:2022tjh}.
At variance with the mid rapidity ALICE data just discussed, 
these measurements are for both forward (right panel) and backward (left panel)
rapidity ranges and, hence, probe rather different values of parton momentum fractions.
Pions produced in the backward direction typically probe nPDFs in the
antishadowing region, $x_b\gtrsim 0.1$, while forward ones 
are mainly sensitive to $x_b\approx 0.01$, i.e., well in the shadowing regime.
This pattern of either a mild enhancement or a more sizable suppression 
due to shadowing is clearly visible in both theoretical estimates obtained with
the two sets of nPDFs combined with vacuum FFs (dashed lines).
Compared to mid rapidity, the theoretical results without invoking final state nuclear
effects now deviate more visibly from data, taking into account the smaller
experimental uncertainties, which, in turn, leaves room for sizable
medium modifications of up to 30\%.
We also note that the pion contribution from quark fragmentation 
becomes quickly comparable in size with pions stemming from gluons
as $p_T$ increases.

Before we come back to these data in Section III to discuss to
what extent the inclusion of nuclear effects in the fragmentation
alters the so far discussed results, we briefly introduce
the second key-observable, namely data on the nuclear SIDIS 
process $eA\to hX$.
Here, it is customary to present the effects of the nuclear medium
as the double ratio $R^h_A$ of the measured hadron multiplicity 
$N^h(\nu,Q^2,z,p_T^2)/N^e(\nu,Q^2)$ in $eA$ to the 
corresponding multiplicity obtained in $eD$ scattering:
\begin{equation}
\label{eq:mult}
R^h_A(\nu,Q^2,z,p_T^2)=\frac{\frac{N^h(\nu,Q^2,z,p_T^2)}{N^e(\nu,Q^2)}\Big|_A}
{\frac{N^h(\nu,Q^2,z,p_T^2)}{N^e(\nu,Q^2)}\Big|_D}\;,
\end{equation}
where $N^h(\nu,Q^2,z,p_T^2)$ is the number of observed hadrons $h$ with 
a given $z$ and $p_T$ in SIDIS and $N^e(\nu,Q^2)$ denotes 
the number of inclusive leptons in DIS at the same kinematics 
defined by $\nu$ and $Q^2$. 

This more involved quantity, often called hadron attenuation, is chosen in order to minimize
medium effects which, in a factorized QCD approach similar to Eq.~(\ref{eq:xsec}),
belong to the initial state nPDFs.
The cancellation would be exact only at leading order (LO) accuracy and if all
initial state nuclear effects could be represented by a single multiplicative factor,
modifying each quark flavor of the nPDFs in the same way.
Nevertheless, this idea behind the double ratio (\ref{eq:mult}) still works remarkably
well even if higher order QCD corrections and more sophisticated 
flavor dependent nuclear effects are accounted for.

In Fig.~\ref{fig2} we show a selection of recent 
measurements of the hadron attenuation $R^{\pi ^+}_A$ (left panel) and $R^{\pi ^-}_A$ (right panel) 
for positively and negatively charged pions, respectively, and different nuclei C, Fe, and Pb
by the CLAS collaboration \cite{CLAS:2021jhm}.
The different bins shown in Fig.~\ref{fig2} correspond to various values 
of the photon's virtuality squared $Q^2$ and transferred energies $\nu$,
which can be translated into different ranges of the parton momentum fraction $x$. 

Likewise, Fig.~\ref{fig3} shows the corresponding nuclear modification factors
$R^{\pi^+}_A$, $R^{\pi^-}_A$, and $R^{\pi^0}_A$ for charged and neutral pions
in eA SIDIS on different noble gas targets (He, Ne, Kr, Xe) in bins of $z$, $x$, and $Q^2$.
These older measurements from HERMES \cite{HERMES:2009uge} are not only taken on different nuclei but also
cover a slightly different kinematics than CLAS. We note that the HERMES data
have been already included in our original analysis of nFFs \cite{Sassot:2009sh}.

As already anticipated, nuclear effects from the initial state, 
as modeled by nPDFs, largely cancel in theoretical estimates of
the nuclear modification factors for SIDIS multiplicities in Eq.~\ref{eq:mult}
and yield results close to unity. 
As can be inferred from the dashed lines
in Figs.~\ref{fig2} and \ref{fig3}, this cancellation 
takes place across all kinematic bins and is also fairly
independent of the atomic mass number $A$.
On the contrary, the data exhibit a sizable suppression
which increases both with $z$ and the nuclear size.
The recent CLAS results shown in Fig.~\ref{fig2} 
nicely confirm and complement the strong attenuation 
of pion yields in SIDIS in a nuclear medium  
first reported by the HERMES collaboration.

Since SIDIS multiplicities at fixed target energies are strongly
dominated by quark fragmentation processes,
the experimental findings by HERMES and CLAS, summarized in
Figs.~\ref{fig2} and \ref{fig3}, already hint towards
a possible explanation through a matching suppression
of quark FFs in a nuclear medium that also increases with 
$z$ and $A$. 
To see whether such a picture can be accommodated 
along with the previously discussed
experimental results from pA, that mainly probe 
possible medium modifications of the gluon FF, 
and, in addition, provides an improved global description of 
both pA and eA data, will be discussed in detail next.

\section{Global Analysis of nFFs revisited}
%
In the following, we pursue the idea of modifying the 
hadronization process in the presence of a nuclear medium
by replacing the vacuum FFs, used so far in obtaining the results
discussed in Sec.~II, by a set of effective nuclear FFs.
To this end, we will closely follow the approach taken in
our original analysis of nFFs \cite{Sassot:2009sh}
and, hence, only briefly recall and summarize the main ideas
and technical steps.
Next, we will determine the optimum amount and shape
of the nuclear modifications to quark and gluon FFs from a global
QCD analysis based on the data presented in Sec.~II, 
followed by a discussion of the results and potential
shortcomings of nFFs.

The main idea is to parametrize the nuclear $A$ dependence of the nFFs in a 
convolutional approach \cite{Sassot:2009sh}, i.e., by setting
\begin{equation}
D_{c/A}^h(z,\mu_0) =
\int_z^1 \frac{dy}{y}\,\, W_c^h(y,A,\mu_0)\,\,D_c^h\left(\frac{z}{y},\mu_0\right)\;.
\label{eq:conv-ansatz}
\end{equation}
Equation (\ref{eq:conv-ansatz}) relates the nFFs $D_{c/A}^h$ of parton flavor $c$
to the corresponding vacuum FFs, in our case taken from the BDSS21 fit \cite{Borsa:2021ran}, 
at some initial scale $\mu_0$ through a weight function $W_c^h$
with the most economical choice of additional fit parameters, see below.
Since sufficient data are only available for identified pions, we have to limit ourselves to
the case $h=\pi$, where the FFs for $\pi^+$ and $\pi^-$ are related by the usual charge 
symmetry assumptions \cite{Borsa:2021ran}.

The nFFs at relevant energy scales $\mu>\mu_0$, with $\mu\simeq p_T$ or $Q$ for pA and eA data,
respectively, are then obtained by applying the standard timelike evolution equations \cite{ref:evol}
at NLO accuracy. 
Since we assume that the nFFs effectively obey final state factorization, but now 
in an $A$-dependent way, the measured observables can be computed in the usual fashion 
with the standard hard scattering partonic cross sections, just replacing the vacuum
FFs in Eq.~(\ref{eq:xsec}) by the medium modified ones.

Contrary to our original analysis \cite{Sassot:2009sh}, the now much more precise 
data allow us to introduce different weights for the favored (valence) and unfavored (sea) quark flavors
in Eq.~(\ref{eq:conv-ansatz}).
Hence, we adopt the following ansatz for positively charged pions:
\begin{eqnarray}
W_v(y,A,\mu_0)&=&n_v \, y^{\alpha_v} (1-y)^{\beta_v} +n'_v\,\delta(1-\epsilon_v-y), \nonumber \\ 
W_s(y,A,\mu_0)&=&n_s \, y^{\alpha_s} (1-y)^{\beta_s} +n'_s\,\delta(1-\epsilon_s-y),\\
\,W_g(y,A,\mu_0)&=&n_g \, y^{\alpha_g} (1-y)^{\beta_g} +n'_g\,\delta(1-\epsilon_g-y), \nonumber
\label{eq:flexi}
\end{eqnarray}
where the subscripts $v$, $s$, and $g$ denote the weight factors for 
valence flavors, sea quarks, and gluons, respectively, and $\delta$ is the usual Dirac delta distribution.
The nuclear $A$ dependence of the coefficients $\xi_i=\{n_i,n_i',\alpha_i,\beta_i,\epsilon_i\}$ 
with $i=v,s,g$ in Eq.~(\ref{eq:flexi}) can be implemented by continuous functions
\begin{equation}
\label{eq:adep}
\xi_i(A)=\lambda_{\xi_i}+\gamma_{\xi_i}\,A^{\delta_{\xi_i}}\,.
\end{equation}
A more complicated functional dependence on the atomic mass number $A$ 
does not improve the quality of the fit. In practice, some of the $\delta_{\xi_i}$ 
often take very similar values in the fit, and
some other parameters usually end up very close to unity or zero 
as is required by the vanishing of nuclear effects in the limit $A \to 1$.
In order to reduce the number of fit parameters, we thus set 
$\delta_{n'_v}=\delta_{\epsilon_v}$, $\delta_{n_v}=\delta_{\alpha_v}=\delta_{\beta_v}$,
$\lambda_{n'_i}=1$, $\lambda_{\epsilon_i}=0$, $\lambda_{n_i}=0$, \ldots; see 
Table~\ref{tab:flexi} below.
In total, 24 free parameters need to be determined by our global fit
to the available 123 pA and 694 SIDIS data points.

We perform our analysis for two different sets of nPDFs,
nNNPDF3.0 \cite{AbdulKhalek:2022fyi} and TUJU21 \cite{Helenius:2021tof},
in order to estimate the dependence of the extracted nFFs on 
what we assume about initial state nuclear effects.
As was already pointed out in Sec.~II, in some cases 
the current uncertainties in nPDFs
can be as large as the expected size of some final state 
nuclear effects, see Fig.~\ref{fig1}.

The results for the nuclear modification factors in pA and eA, computed with
our new estimates for final state nuclear effects, are represented by the solid
lines in Figs.~\ref{fig1}-\ref{fig3}.
The shaded bands around the results based on the nNNPDF3.0 set of nPDFs 
(solid red lines) correspond to estimates 
of the uncertainties of our nFFs at the $68\%$ confidence level (C.L.)
due to the experimental errors of the fitted data.
These estimates are obtained by constructing and fitting a set of 600 Monte Carlo
replicas of the available data and then computing the average (best fit)
and variance of the resulting nFFs (shaded band). 
We note, that the corresponding uncertainty bands for the fit based on the
TUJU21 set of nPDFs are very similar in size and shape. Hence,
for the sake of clarity, we refrain from showing them 
as well in Figs.~\ref{fig1}-\ref{fig3}.

As can be inferred from Figs.~\ref{fig2} and \ref{fig3}, both
fits of nFFs reproduce very well the main features
of the rather large hadron attenuation found in the
nuclear modification factors for SIDIS multiplicities
by CLAS and HERMES.
More importantly, this happens not only for the measured
$z$ and $A$ dependence of the data, which is more directly
related to the extracted nFFs, but also their 
$x$ and $Q^2$ dependence is well reproduced. The latter
observation is much less obvious due to the 
convoluted interplay with the $x$ and $Q^2$ dependence of the nPDFs
and is best seen in Fig.~\ref{fig3} as the HERMES data cover
a larger kinematic range in $x$ and $Q^2$ than CLAS.
%
\begin{table}[t!]
\caption{\label{tab:exppiontab} Data sets, normalizations $\mathcal{N}_i$, and the partial 
and total $\chi^2$ values obtained in our fit for both choices of nPDFs
(TUJU21 and nNNPDF3.0).}
\begin{ruledtabular}
\begin{tabular}{llccccc}
experiment& data & data  & \multicolumn{2}{c}{{\footnotesize TUJU21 }} & \multicolumn{2}{c}{{\footnotesize nNNPDF3.0 }} \\
          & type & fitted&                  $\mathcal{N}_i$ & $\chi^2$ & $\mathcal{N}_i$ & $\chi^2$ \\ \hline
HERMES \hfill \cite{HERMES:2009uge} & He - $\pi^+$        & 25  &   1.010  & 26.8 & 1.008 & 30.6 \\
                                    & He - $\pi^-$        & 25  &   1.017  & 17.9 & 1.008 & 17.8 \\
                                    & He - $\pi^0$        & 25  &   1.026  & 21.9 & 1.023 & 22.3 \\
                                    & Ne - $\pi^+$        & 25  &   1.024  & 44.8 & 1.018 & 43.1 \\
                                    & Ne - $\pi^-$        & 25  &   1.023  & 36.4 & 1.011 & 32.7 \\
                                    & Ne - $\pi^0$        & 25  &   1.035  & 42.3 & 1.027 & 40.2 \\
                                    & Kr - $\pi^+$        & 25  &   1.035  & 81.8 & 1.032 & 76.4 \\
                                    & Kr - $\pi^-$        & 25  &   1.021  & 66.9 & 1.014 & 57.4 \\
                                    & Kr - $\pi^0$        & 25  &   1.035  & 68.8 & 1.031 & 63.6 \\
                                    & Xe - $\pi^+$        & 25  &   1.049  & 122.4 & 1.042 & 108.8 \\
                                    & Xe - $\pi^-$        & 25  &   1.027  & 75.1 & 1.028 & 76.4 \\
                                    & Xe - $\pi^0$        & 25  &   1.039  & 117.9 & 1.033 & 117.9 \vspace{1mm} \\
CLAS \hfill \cite{CLAS:2021jhm}     & C - $\pi^+$         & 72  &   0.965  & 305.0 & 0.956 & 124.9 \\
                                    & C - $\pi^-$         & 60  &   0.991  & 194.7 & 0.974 & 114.2 \\
                                    & Fe - $\pi^+$        & 72  &   0.910  & 400.3 & 0.912 & 288.3 \\
                                    & Fe - $\pi^-$        & 62  &   0.962  & 240.4 & 0.926 & 272.4 \\
                                    & Pb - $\pi^+$        & 72  &   0.931  & 230.6 & 0.960 & 166.9 \\
                                    & Pb - $\pi^-$        & 56  &   0.988  & 211.6 & 0.973 & 275.6 \\ \hline
{\bf SIDIS eA data }                &                     & 694 &          &2305.6 &       & 1929.5  \\\hline
PHENIX \hfill \cite{PHENIX:2006mhb} & dAu - $\pi^0$    & 10  &   1.006  & 12.9 & 1.003 &  3.5 \\
STAR \hfill \cite{STAR:2003oii}     & dAu - $\pi^0$    & 12  &   1.002  & 14.1 & 1.001 &  8.8 \\
ALICE \hfill \cite{ALICE:2016dei}   & pPb - $\pi^0$    & 24  &   0.987  & 16.9 & 1.008 &  7.4 \\
\hfill  \cite{ALICE:2016dei}        & pPb - $\pi^\pm$  & 35  &   0.989  & 60.6 & 1.009 &  9.4 \\
LHCb \hfill bwd \cite{LHCb:2022tjh} & pPb - $\pi^0$    & 21  &   1.045  & 110.3 & 1.023 & 35.7 \\
     \hfill fwd \;\;\;\;\;\,\,      & pPb - $\pi^0$    & 21  &   0.913  & 212.2 & 0.952 & 81.6 \\ \hline
 {\bf p(d)A data }                     &                   & 123 &          & 427.0 &       & 146.4 \\
 \hline\hline
{\bf TOTAL:} & & 817 &  & 2732.6 & & 2075.9 \\
\end{tabular}
\end{ruledtabular}
\end{table} 

The large nuclear modifications of the quark FFs
required to fit the SIDIS data 
do not spoil the description of the
nuclear modification ratios in hadroproduction from the LHC though.
On the contrary, the agreement with the mid rapidity data from ALICE,
shown in the upper panels of Fig.~\ref{fig1},
is also significantly better for our fit of final state nuclear effects
based on the nNNPDF3.0 set of nPDFs (solid red lines).
Since the low $p_T$ region at mid rapidity predominantly probes 
gluon initiated hadronization processes, this improvement is readily 
explained by an enhancement of the gluon nFF due to medium effects.
The corresponding results adopting the nFFs based on the TUJU21 nPDFs 
as baseline (blue solid lines), compares less favorably to the ALICE data.
In this case, a better description of the data would require less suppression 
for the quark nFFs and less nuclear enhancement of the gluon nFF, 
both of which are disfavored in the global fit, however, 
by the SIDIS and LHCb data, respectively.

The LHCb data taken at forward and backward rapidity, see the lower panels of Fig.~\ref{fig1},
help to discriminate different regions of parton momentum fraction in the nPDFs,
more precisely, the amount of shadowing and anti-shadowing found 
in the respective $x$ regimes.
Apparently, the forward data (lower right panel), prefer the larger shadowing 
present in the nNNPDF3.0 set of nPDFs compared to the more moderate 
one found in TUJU21 to counteract the enhancement of the gluon nFF 
needed at mid rapidity. 
The general agreement with the LHCb data at backward rapidity is again 
found to be better with the nNNPDF3.0 set of nPDFs and, here, the increased
gluon fragmentation goes into the right direction.
Cleary, in both regions the description of the LHCb data leaves a lot
of room for improvements though.
Due to the intricate interplay between initial and final state medium effects
at play here, a more detailed study would strongly benefit
from a combined global analysis of nPDFs and nFFs at some point in the future.
Such a procedure has already been explored for vacuum FFs and PDFs by means
of an iterative approach in Ref. \cite{Borsa:2017vwy}.
Within the current uncertainties of nPDFs it would be certainly
possible to reshuffle the amount of shadowing and anti-shadowing against
the pattern of suppression and enhancement in the quark and gluon
nFFs found in our analysis. Most likely, this would lead to further 
improvements in the theoretical description of hadroproduction data 
in pA collisions in all rapidity regimes.

In Table~\ref{tab:exppiontab} we summarize the data sets 
used in our NLO global QCD analysis of parton-to-pion nFFs
by stating the individual contributions of each data set
to the total $\chi^2$ for both choices of baseline sets of nPDFs.
We also give the normalization shifts $\mathcal{N}_i$ for each data set.
These are not fitted but computed analytically as defined in 
Eq.~(6) of Ref.~\cite{deFlorian:2014xna}. They are all very reasonable and
turn out to be close to unity, i.e., in the range of typical experimental
normalization uncertainties.

We note, that our fit also includes older hadroproduction data
from the two RHIC experiments PHENIX \cite{PHENIX:2006mhb} 
and STAR \cite{STAR:2003oii} taken in dAu collisions at
a much lower c.m.s.\ energy than the LHC results.
At the time of our original fit \cite{Sassot:2009sh}
these were the only measurements available on 
nuclear modification factors apart from the 
hadron attenuation data from HERMES.
As can be seen from Tab.~\ref{tab:exppiontab}, both
sets are still well described also by our new fit.

The figures of merit $\chi^2_\text{TUJU21}$ and $\chi^2_\text{nNNPDF3.0}$ 
resulting from the respective $\chi^2$-optimizations of the 24 free parameters
defined in Eqs.~(\ref{eq:conv-ansatz}) - (\ref{eq:adep})
only include the quoted experimental errors of the data sets listed
in Tab.~\ref{tab:exppiontab}. They are not inflated by the various
potential sources of theoretical ambiguities stemming from the
adopted sets of PDFs, nPDFs, and vacuum FFs or the choice of
factorization and renormalization scales in Eq.~(\ref{eq:xsec}).
Even though some of these uncertainties tend to cancel in the 
relevant nuclear modification ratios, 
the residual uncertainties are nevertheless expected to be 
in total larger than the experimental errors, explaining the
rather sizable $\chi^2$ per data point of our fit.
We also wish to mention, that, for simplicity,
we have computed the nuclear SIDIS multiplicity ratios 
at the center of each experimental bin and disregard any possible
bin size effects, which can be particularly large for the 
larger $Q^2$ bins of the CLAS data.

Clearly, the quoted $\chi^2_\text{TUJU21}$ and $\chi^2_\text{nNNPDF3.0}$ numbers 
should not be judged as in a textbook parameter fitting exercise to a perfect theory, 
but rather as a comparative tool in a hypothesis testing context.
Nevertheless, the results obtained with the nNNPDF3.0 set of nPDFs are
noticeably better than those using the TUJU21 set. 
Most significantly is the difference in the total $\chi^2$ for the
hadroproduction data in p(d)A collisions, as can be inferred from
Tab.~\ref{tab:exppiontab}, but also the eA SIDIS multiplicity
ratios are much better described by the fit using nNNPDF3.0 as the baseline.
For completeness, the optimum fit parameters for the weight
functions defined in Eqs.~(\ref{eq:conv-ansatz}) - (\ref{eq:adep})
for our new set of nFFs based on nNNPDF3.0 nPDFs can
be found in Table \ref{tab:flexi}.
%
%
\begin{table}[ht!]
\caption{\label{tab:flexi}Optimum set of parameters for our 
nFFs based on the nNNPDF3.0 set of nPDFs, see
Eqs.~(\ref{eq:conv-ansatz}) - (\ref{eq:adep}),
at the input scale $\mu_0=1\,\mathrm{GeV}$ of our fit.}
\begin{ruledtabular}
\begin{tabular}{cccc}
   & $\lambda_i$ & $\gamma_i$ & $\delta_i$  \\ \hline
$n'_v$        &      1 & -0.0801 & 0.3999  \\ 
$\epsilon_v $ &      0 & 0.0018  & 0.3999  \\
$n_v$         &      0 & 0.1342  & 0.0166  \\ 
$\alpha_v$    &-5.1826 & 6.0180  & 0.0166  \\  
$\beta_v $    &-5.4788 & 5.3592  & 0.0166  \\ \hline
$n'_s$        &      1 & -0.0481 & 0.3999  \\ 
$\epsilon_s $ &      0 &  0.0018 & 0.3999  \\
$n_s$         &      0 &  0.0739 & 0.2079  \\ 
$\alpha_s$    &-5.4510 & 4.0912  & 0.2079  \\  
$\beta_s $    & 1.2741 & 3.5697  & 0.2079  \\ \hline
$n'_g$        &      1 & 0.0574  & 0.3999  \\ 
$\epsilon_g $ &      0 &     0   &      0  \\
$n_g$         &      0 & -0.0141 & 0.6157  \\ 
$\alpha_g$    & 8.0685 & 0.0352  & 0.6157  \\  
$\beta_g $    & 0.0009 & 0.0064  & 0.6157  \\
\end{tabular}
\end{ruledtabular}
\end{table}

\begin{figure*}[th!]
\epsfig{figure=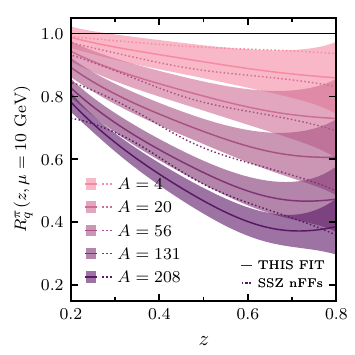,width=0.35\textwidth}
\epsfig{figure=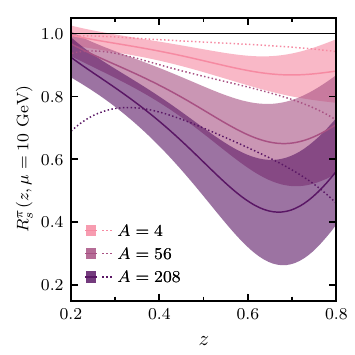,width=0.35\textwidth}
\vspace*{-0.2cm}
\caption{The ratio $R^h_{i/A}(z,\mu)$ defined in Eq.~(\ref{eq:dratio}) at $\mu=10\,\mathrm{GeV}$
for valence (left-hand-side) and
sea quarks (right-hand-side) as a function of $z$ for various nuclei $A$.
The uncertainty bands at $68\%$ C.L.\ are obtained from the Monte Carlo replicas. For comparison we include the ratios obtained in ref. \cite{Sassot:2009sh} as dotted lines.}\label{fig5}
\vspace*{0.2cm}
\epsfig{figure=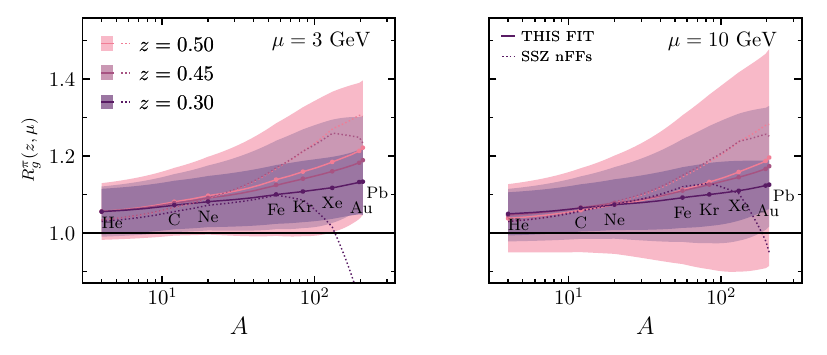,width=0.8\textwidth}
\vspace*{-0.2cm}
\caption{Similar as in Fig.~\ref{fig5} but now for the gluon at
scale $\mu=3\,\mathrm{GeV}$ (left-hand-side) and $\mu=10\,\mathrm{GeV}$ (right-hand-side)
and as a function of $A$ for three different values of $z$. Also shown are the respective
uncertainty bands at $68\%$ C.L. and the ratios from ref. \cite{Sassot:2009sh} as dotted lines.}\label{fig6}
\end{figure*}
In order to visualize the resulting medium modifications of 
the FFs, we study the ratio
\begin{equation}
\label{eq:dratio}
R^h_{i/A}(z,\mu) \equiv \frac{D^h_{i/A}(z,\mu)}{D_i^h(z,\mu)}
\end{equation}
between the nuclear and the vacuum FF for different values
of $z$, the atomic mass number $A$, and scale $\mu$.
We compute (\ref{eq:dratio}) for the set of nFFs $D^h_{i/A}(z,\mu)$
based on the nPDFs from nNNPDF3.0, and the $D_i^h(z,\mu)$
are always taken from the BDSS21 fit \cite{Borsa:2021ran}.

Figure~\ref{fig5} shows the ratios $R^h_{i/A}(z,\mu)$ for 
valence (left-hand-side) and sea quarks (right-hand-side) 
as a function of $z$ for a scale $\mu=10\,\mathrm{GeV}$
and various atomic mass numbers $A$.
Since the quark nFFs are mainly determined by the large
nuclear modifications of the SIDIS multiplicities shown
in Figs.~\ref{fig2} and \ref{fig3}, it is not surprising
that the nFF-to-FF ratios (\ref{eq:dratio}) exactly mimic this
behavior. Thus we observe a strong suppression of
quark fragmentation in a nuclear medium that increases further
with both $z$ and $A$.
The corresponding uncertainty bands exhibit only a moderate dependence on $z$
but, as expected, their width increases at both endpoints of the range in $z$ covered by the 
HERMES and CLAS SIDIS data.
As a function of nuclear size, larger nuclei tend to be better constrained
than lighter ones. This observation could be related to the fact that the results 
for Pb nuclei are constrained by both SIDIS and pPb data.

Figure 4 shows also the results of ref. \cite{Sassot:2009sh} for the quark nuclear FFs as dotted
lines. In the case of the valence quark FFs in the left hand side panel, the new
FFs show a slightly larger suppression than the older set, but they have a similar
trend as a function of $z$ and they agree within uncertainties. For the sea quarks
in the right hand size panel, the differences are larger, however one should keep
in mind that in ref. \cite{Sassot:2009sh} sea quark FFs were assumed to be the same as those for valence quarks, since no significant improvement was found discriminating them.
The new set of data, specifically the very precise CLAS data, now favors different
shapes for the valence and the sea quark fragmentation.

For the nuclear modification factors measured in hadro\-production, 
see Fig.~\ref{fig1}, the most relevant quantity is now the 
amount of nuclear effects on the gluon FF.
Rather than showing the gluon nFF-to-FF ratio (\ref{eq:dratio}) as a function of $z$ 
as in Fig.~\ref{fig5}, it is perhaps more interesting to focus on the $A$ dependence
for typical scales of the order $\mu\simeq p_T$.
This is because the kinematics is such, that the data predominantly probe 
only a rather narrow range of $z$ centered around $z\simeq 0.45$ and $z\simeq 0.50$
for LHC and RHIC data, respectively; see Ref.~\cite{Sassot:2010bh}.
Hence, in Fig.~\ref{fig6} we show instead the gluon nFF-to-FF ratio as a function of $A$ for
two representative values of $\mu=p_T$, 
$\mu=3\,\mathrm{GeV}$ (left-hand-side) and $\mu=10\,\mathrm{GeV}$ (right-hand-side),
roughly corresponding to the $p_T$ range covered by the data shown in Fig.~\ref{fig1}.
The bands illustrate the uncertainties at $68\%$ C.L.\ for three different
values of $z$ (0.3, 0.45, and 0.5) as determined from our Monte Carlo replicas.

Contrary to the case of quark fragmentation, the main effect for gluons is an
enhancement in the probability to produce hadrons in a nuclear medium.
The increase is less pronounced than the suppression found for quark nFFs though,
and also the dependence on $A$ is more moderate.
The gluon nFF-to-FF ratio being larger than unity is mainly a result of improving the overall 
agreement with the LHC pPb data shown in Fig.~\ref{fig1} as is best seen by
comparing the dashed to solid red lines. At larger values of $p_T$ though
($\approx10\,\mathrm{GeV}$ at mid rapidity and $\approx5\,\mathrm{GeV}$ at forward or
backward rapidity), the strong suppression of the quark nFFs starts to dominate 
final state effects also in hadroproduction.
Somewhat counterintuitive is the finding that the uncertainties increase for larger nuclei despite
the fact that the data mainly constraining the gluon nFF are from hadroproduction
in pPb and dAu collisions. This is perhaps an artifact of the gluon FF itself being numerically
very small at the relevant values of $z$ around 0.45 that are predominantly probed in hadroproduction.

Compared to the previous extraction of gluon nFFs in ref. \cite{Sassot:2009sh}, the new set shows a
similar nuclear size dependence for $z=0.45-0.50$, where the gluons are better
constrained. For $z=0.3$ the differences are very significative, most likely because
of the comparative lack of constrains for the gluon at lower $z$ through the scale
dependence of the other distributions in the previous fit.

\begin{figure*}[t!]
\hspace*{-1.4cm}
\epsfig{figure=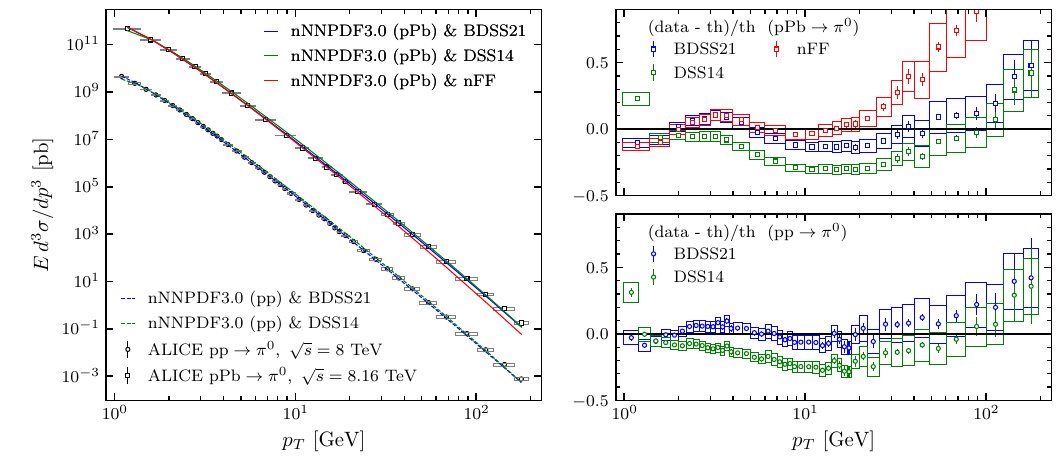,width=0.8\textwidth}
\vspace*{-0.2cm}
\caption{The pp and pA cross section for hadroproduction of neutral pions at $\sqrt{S}\simeq 8\,\mathrm{TeV}$
from ALICE \cite{ALICE:2021est} compared to theoretical estimates with different sets of FFs and nFFs (left-hand-side) 
and the corresponding ratios for ''(data-theory)/theory'' (right-hand-side), see text.}\label{fig7}
\end{figure*}
Finally, we turn to the latest piece of experimental data on $R_{pPb}$ for neutral pions 
at mid rapidity from ALICE obtained at a higher energy of $\sqrt{S}=8.16\,\mathrm{TeV}$ and also extending the
range of $p_T$ to about $200\,\mathrm{GeV}$ \cite{ALICE:2021est}.
As we have mentioned before, these data were not used in the present analysis as 
the corresponding pp data were not yet included in the fit of the BDSS21 vacuum FFs, which we use
as a baseline to quantify medium induced effects. 
We want to elaborate on this a bit further and also wish to point out an 
important aspect when fitting only to ratios of cross sections that should not be overlooked.

ALICE reports that the measured $R_{pPb}$ is consistent with unity for $p_T>10\,\mathrm{GeV}$ 
\cite{ALICE:2021est}, which they interpret as an indication for the absence of final state
nuclear effects at high $p_T$, and which agrees well with theoretical
estimates based on recent sets of nPDFs when combined with the DSS14 vacuum FFs \cite{deFlorian:2014xna}.
However, looking only at cross section ratios can hide a lot of theoretical 
issues, as we illustrate in Fig.~\ref{fig7}.
As can be seen, neither the pp nor the pA cross section are well described by using the DSS14 vacuum FFs. 
Both results deviate from data significantly, but roughly by the same amount and in the
same direction for $p_T\gtrsim 10\,\mathrm{GeV}$, such that the reported agreement of
the ratio $R_{pPb}$ with data is not more than a mere coincidence.

This highlights the importance of {\em first} having adequate estimates 
for the pp baseline cross section before drawing any conclusions about
the relevance or absence of nuclear modifications.
As can be also inferred from Fig.~\ref{fig7}, the theoretical estimates
improve considerably when adopting the more recent BDSS21 set of FFs (blue boxes)
but deviations in shape are still outside of the range of experimental errors for
basically all values of $p_T$. For that reason, we refrained from using 
the most recent data from ALICE in our present global analysis, despite of
their uniqueness in terms of the very much extended reach in $p_T$.

A detailed analysis of these data unfortunately has to await a reanalysis of 
vacuum FFs to first arrive at the best possible theoretical estimate for the pp
baseline cross section. However, to give at least some idea of how our new
set of nFFs performs, with all the caveats mentioned above,
we also show in Fig.~\ref{fig7} the pPb cross section based on our fit (red boxes). 
We find that 
up to $p_T\lesssim 22\,\mathrm{GeV}$ the inclusion of final state nuclear
effects significantly improves its theoretical description. 
In this range, the $\chi^2$ per data point for the pPb cross section
drops from 42.5 (with DSS14) to 5.9 (with BDSS21) to 2.0 with nFFs.

For larger $p_T$ though, the agreement quickly deteriorates due to the
strong suppression of the quark nFFs as can be seen as well in Fig.~\ref{fig7}.
It could very well be, that this points to 
limitations of the effective factorization assumed in modeling the nFFs,
but clarifying this question would first require a combined 
global analysis of nPDFs and nFFs to fully exploit the intricate
interplay of shadowing and antishadowing in nPDFs on the one hand and
the patterns of suppression and enhancement found in quark and gluon nFFs 
on the other hand.

Our sets of nFFs are available in LHAPDF format \cite{buckley2015lhapdf6} from the authors by request.

\section{Conclusions and Outlook}
%
We checked the viability of modeling final state nuclear effects on
pion yields in pA and eA collisions through a set of effective nuclear
fragmentation functions in the light of a plethora of new and precise
experimental data.

Based on a global QCD analysis at next-to-leading order accuracy,
we showed that the hadronization process in such nuclear environments 
can be simultaneously described by a single set of nuclear
fragmentation functions that rely on an effective, nuclear size dependent
final state factorization.
The modifications of vacuum fragmentation functions are described
in a convolutional approach with only a limited number of fit parameters
used to define a set of weight functions.

The eA SIDIS and pA hadroproduction data leads to a significant suppression 
of quark fragmentation, that increases both with momentum fraction and nuclear size, 
and a more moderate enhancement of gluon fragmentation, respectively.
These findings align well with our original analysis of medium modified
fragmentation functions that was based on a much more limited set of 
measurements, covering also a significantly smaller kinematic range.
Overall, the inclusion of final state nuclear effects significantly
improves the global description of eA and pA data as compared to
theoretical computations based solely on modifications of the
initial state parton densities combined with unaltered vacuum fragmentation functions.

To fully explore the interplay of nuclear effects on parton distribution and
fragmentation functions in an unbiased way, a combined global analysis 
of these nonperturbative quantities must be set up and performed in the future. 
This will ultimately help to unambiguously reveal potential limitations of the
assumed effective factorization of final state nuclear effects,
for which, perhaps, the latest large transverse momentum data already provide
a first hint. Future experimental results from the LHC and, eventually, from
a first Electron-Ion Collider in about a decade from now,
will also add to our understanding of nuclear effects on both 
parton distribution and fragmentation functions.

\section*{Acknowledgments}
%
R.M.\ wishes to thank Hampton University and JLab for hospitality during the completion of this work.
We warmly acknowledge Alberto Accardi for useful comments and suggestions 
and Ilkka Helenius for providing us with the TUJU21 nPDF set. 
This work was supported in part by CONICET, ANPCyT, UBACyT, 
and the Bundesministerium für Bildung und Forschung (BMBF) under grant no.\ 05P21VTCAA. 
%

\end{document}